\documentstyle[12pt,epsfig,
 ]{article}
\begin{document}
\begin{titlepage}
\vspace{-2mm} \rightline{hep-th/0204116} \vskip 1cm
\centerline{\Large {\bf  Time-dependent driven anharmonic
oscillator}} \vspace{1.5mm} \centerline{\Large {\bf and the
adiabaticity  }} \vspace{1.5mm}
\vspace{10mm}
\centerline{ Hyeong-Chan Kim$^\dagger$
\footnote{E-mail: \texttt{hckim@phya.yonsei.ac.kr}} and Jae Hyung
Yee$^{\dagger\star}$ \footnote{E-mail:
\texttt{jhyee@phya.yonsei.ac.kr}}}

\centerline{$^\dagger$\it Institute of Basic Science, Yonsei
University, Seoul 120-749, Korea}
\centerline{$^\star$\it Institute of Physics and Applied Physics,
Yonsei University, Seoul 120-749, Korea}
\vskip 5mm \centerline{(\today)}
\setlength{\footnotesep}{0.5\footnotesep}

\vskip 10mm

\begin{abstract}
We  use the Liouville-von Neumann (LvN) approach to study
the dynamics and the adiabaticity of a time-dependent
driven anharmonic oscillator as an example of
non-equilibrium quantum dynamics. We show that the
adiabaticity is minimally broken in the sense that a
gaussian wave packet at the past infinity evolves to
coherent states, however slowly the potential changes, its
coherence factor is order of the coupling. We also show
that the dynamics are governed by an equation of motion
similar to the Kepler motion which satisfies the angular
momentum conservation.
\end{abstract}

\vspace{5mm}

\indent\indent\hspace{4mm}  {\bf Keywords:} harmonic,
anharmonic oscillator, driven, adiabatic, \\
\indent\indent\hspace{4mm} Liouville-von Neumann approach
\vspace{3mm} \indent\indent\hspace{4mm}

{\bf PACS numbers:}

\end{titlepage}

\setcounter{footnote}{0}

\section{Introduction}

The quantization of a time-dependent harmonic oscillator
has been important in many branches of physics such as the
coherent states and the squeezed states~\cite{meystre}, the
quantum statistical properties of
radiation~\cite{louisell}, the information
theory~\cite{wootters}, and the quantum theory out of
thermal equilibrium~\cite{jackiw}. As a more realistic
system anharmonic oscillator has been widely studied for
various purposes. In this direction, considerable efforts
have been devoted to constructing systematic procedures to
improve the variational approximation method of quantum
mechanics and quantum field
theory~\cite{diney,kleinert,lee}. These methods give
appropriate effective potentials which are adequate for
describing time-independent systems. However, if one is
interested in the explicit time evolution of time-dependent
systems, above methods are limited in their uses.

The path integral approach~\cite{benamira} was developed to
study several kinds of time-dependent oscillators with
perturbative external forces. The LvN approach to the
generalized driven oscillator~\cite{kim} was developed and
its geometric phases and wave functions were
discussed~\cite{mhlee}. The LvN approach to the
time-dependent harmonic~\cite{lvn} and anharmonic
oscillators~\cite{spkim} have also been developed. A merit
of LvN approach is that it can be directly applied to both
time-independent and time-dependent quantum systems without
any modification.

In the case of a driven harmonic oscillator, it was
shown~\cite{holstein} that, the transition amplitudes
between different states are zero in the adiabatic limit of
a driving force of the form $F e^{-t^2/\tau^2}$. The
adiabatic assumption, that this phenomena holds for a
general type of quasi-static potentials is usually accepted
but no concrete proof has been given in general situation.
In this paper we test this assumption by checking the
evolution of a gaussian ground state of the time-dependent
driven anharmonic oscillator. We formulate the quantization
of a time-dependent driven anharmonic oscillator with
general time-dependent mass, frequency, and driving force.
The present formulation can also be applicable to the study
of phase transition using the inverted mass
square~\cite{spkim}.

The organization of this paper is as follows. In Sec. II,
the LvN approach is elaborated to be applicable to the
driven anharmonic oscillator and we find the Fock space and
expectation values of some operators. Sec. III is devoted
to study driven anharmonic oscillator system which is
appropriate to analyze the adiabatic assumption. We show
that the initial gaussian states evolve to final coherent
states with its squeezing varies in time by $O(\lambda)$.
In Sec. IV, we consider the weak coupling limit of the
oscillator in Sec. III and we show that the squeezing
becomes time-independent in the adiabatic limit. Therefore,
the adiabatic assumption is broken minimally. In Sec. V, we
summarize our results and give some discussions. We have
two appendices which present some calculational details.

\section{ Time-dependent driven anharmonic oscillator}

The quantization of a time-dependent anharmonic oscillator
using the LvN approach was tried in Ref~\cite{spkim}. It
was shown that their approach is equivalent to the
variational perturbation method~\cite{chang} and the
mean-field, Hartree-Fock method. In this section, we extend
the LvN method to be applicable to the driven anharmonic
oscillator, whose hamiltonian is given by
\begin{eqnarray}
\hat{H}(t)= \frac{\hat{p}^2(t)}{2 m(t)} + \frac{1}{2} m(t)
\omega^2(t) \hat{q}^2(t) - \sqrt{m(t)} F(t) \hat{q}(t)+
\frac{m^2(t) \lambda(t)}{4!}\hat{q}^4(t),
\end{eqnarray}
where we have introduced the mass $m(t)$ in front of the
coupling and the external force for simplicity of the later
equations. The mass dimensions of $f$ and $\lambda$ are 3/2
and 3, respectively.

The LvN approach resorts to find a pair of invariant operators
\begin{eqnarray}
\hat{A} &=& i \left[v^*(t) \hat{p}(t) - m(t) \dot{v}^*(t)
\hat{q}(t)
+ x^*(t) \right] , \\
\hat{A}^\dagger &=&-i \left[v(t) \hat{p}(t) - m(t)
\dot{v}(t) \hat{q}(t) + x(t) \right] , \nonumber
\end{eqnarray}
which satisfy the LvN equation,
\begin{eqnarray}
\frac{d \hat{A}}{dt}&=& \frac{\partial \hat{A}}{\partial
t}-i
   [\hat{A}, \hat{H}] = 0 , \label{lvn} \\
 \frac{d \hat{A}^\dagger}{dt}
   &=& \frac{\partial \hat{A}^\dagger}{\partial t}-i
   [\hat{A}, \hat{H}] = 0 ,  \nonumber
\end{eqnarray}
and the standard commutation relation
\begin{eqnarray}\label{comm:A}
[\hat{A},\hat{A}^\dagger]=1 .
\end{eqnarray}
The commutation relation identifies $\hat{A}$ and
$\hat{A}^\dagger$ as annihilation and creation operators
respectively and it gives a Wronskian condition for $v(t)$:
\begin{eqnarray} \label{comm:v}
\hbar [\dot{\tilde{v}}^*(t) \tilde{v}(t)-
\dot{\tilde{v}}(t)\tilde{v}^*(t)]=i,
\end{eqnarray}
where $v(t) = \tilde{v}(t)/\sqrt{m(t)}$. Later in this
paper, we use the unit which makes $\hbar =1$. The ground
state is the one annihilated by $\hat{A}$ and the $n$~th
number state is obtained by applying $\hat{A}^\dagger$ $n$
times to the vacuum state:
\begin{eqnarray} \label{0:I}
\hat{A}|0\rangle &=& 0, \\
 |n \rangle &=&
\frac{[\hat{A}^\dagger]^n}{\sqrt{n!}}|0\rangle.
\end{eqnarray}
The wave function of the ground state to this order of
approximation is given by
\begin{eqnarray}\label{wf}
\Psi_{0}(q,t) = \left[\frac{1}{2 \pi v^*(t)
v(t)}\right]^{1/4} \exp \left[i \frac{m(t)\dot{v}^*(t)}{2
v^*(t)}q^2- i \frac{x^*(t)}{v^*(t)} q \right].
\end{eqnarray}

The LvN equation~(\ref{lvn}) reduces to the following
operator equations:
\begin{eqnarray}
&&\dot{x}(t) = -F(t) \tilde{v}(t), \label{dx:fv}\\
&&\left[\ddot{\tilde{v}}(t)  + \tilde{\omega}^2(t)
\tilde{v}(t)\right]\hat{q}(t) + \frac{m(t)\lambda(t)
\tilde{v}(t)}{3!} \hat{q}^3(t) = 0, \label{v:qqq}
\end{eqnarray}
with the new frequency $\tilde{\omega}^2(t)
=\omega^2(t)-\frac{1}{ \sqrt{m(t)} } \frac{ d^2 \sqrt{m(t)}
}{d t^2}$. Differentiating Eq.~(\ref{v:qqq}) with respect
to $q(t)$ and taking the vacuum expectation value, which is
equivalent to the variational approximation~\cite{spkim},
one gets
\begin{eqnarray} \label{veq1}
\ddot{\tilde{v}}(t) + \left[\tilde{\omega}^2(t)
+\frac{m(t)\lambda(t) }{2} \langle
0|\hat{q}^2(t)|0\rangle\right] \tilde{v}(t) = 0.
\end{eqnarray}
From the first basic comparison theorem of ordinary
differential equation, one can naturally conclude that the
solution for $\tilde{v}(t)$ should be oscillatory if
$\tilde{\omega}^2$ and $\frac{m(t)\lambda(t) }{2} \langle
0|\hat{q}^2(t)|0\rangle$ is positive definite. The quantum
operators $\hat{q}(t)$ and $\hat{p}(t)$ can be written in
terms of the creation and the annihilation operators as
\begin{eqnarray}
\hat{q}(t)&=& v(t) \hat{A} + v^*(t) \hat{A}^\dagger
+ 2 \Im[v(t) x^*(t)],  \label{q:A}  \\
\hat{p}(t)&=& m(t)\left\{\dot{v}(t)\hat{A} +
\dot{v}^*(t)\hat{A}^\dagger + 2 \Im[\dot{v}(t)
x^*(t)]\right\},  \label{p:A} \nonumber
\end{eqnarray}
where $\Im$ means the imaginary part and
\begin{eqnarray} \label{x:fv}
x(t) = - \int _{-\infty}^t F(t) \tilde{v}(t) dt .
\end{eqnarray}

The vacuum expectation values of operators are then given
by
\begin{eqnarray}\label{0:op}
\langle0|\hat{q}(t)|0 \rangle &=&  2\Im[v(t)
x^*(t)]  , \\
\langle 0|\hat{p}(t)|0 \rangle &=& 2m(t)\Im[\dot{v}(t)
x^*(t)] \nonumber , \\
\langle 0|\hat{q}^2(t)|0 \rangle &=& |v(t)|^2 +
4\left|\Im[v(t)
x^*(t)]\right|^2,  \nonumber \\
\langle 0|\hat{p}^2(t)|0 \rangle &=& m^2(t)|\dot{v}(t)|^2 +
4m^2(t) \left|\Im[\dot{v}(t) x^*(t)]\right|^2 . \nonumber
\end{eqnarray}
The quantum mechanical dynamics of the forced anharmonic
oscillator system are then contained in one classical
equation of motion ~(\ref{veq1}), which becomes
\begin{eqnarray} \label{ddv:o}
\ddot{\tilde{v}}(t) + \omega_{eff}^2(t) \tilde{v}(t) = 0
\end{eqnarray}
where
\begin{eqnarray}\label{oeff:v}
\omega_{eff}^2(t)=\tilde{\omega}^2(t) +\frac{\lambda(t)
}{2}\left[|\tilde{v}(t)|^2+ \left|\tilde{v}(t)
x^*(t)-\tilde{v}^*(t) x(t)\right|^2 \right].
\end{eqnarray}
In summary, we have reduced the problem of quantization of
driven anharmonic oscillator into the problem of solving
one classical differential equation~(\ref{ddv:o}). All
information of the dynamics are encoded in a single
function $\tilde{v}(t)$. Once the explicit form of
$\tilde{v}(t)$ is known, we have time evolution of the
quantum operators $\hat{p}(t)$ and $\hat{q}(t)$, and time
evolution of any other physical quantities are derived from
these operators and the states~(\ref{0:I}).

\section{Anharmonic oscillator with a
linearly growing external force}

In this section, we present the evolution of the driven
anharmonic oscillators with constant and linearly
increasing forces. Let us first analyze the case of
time-independent driven oscillator. We assume
$\tilde{\omega}(t)$ and $F(t)$ to be time-independent. The
mass $m(t)$ and the frequency $\omega(t)$ may be dependent
on time, whose time dependency leads to no trouble as long
as they are determined to give time-independent
$\tilde{\omega}$.

An exact solution of Eq.~(\ref{ddv:o}) can be found by
assuming
\begin{eqnarray}
x(t) = - F r \exp (-i \Omega t),~~ \tilde{v}_0(t) = -i
\Omega r \exp (-i \Omega t) ,
\end{eqnarray}
where $F$ and $r$ are constants. The value of $r$ should be
given from the Wronskian formula~(\ref{comm:v}), and the
resulting solution is
\begin{eqnarray} \label{v00:t}
v_0(t)= \frac{1}{\sqrt{2\Omega}} e^{-i \Omega t}.
\end{eqnarray}
Substituting~(\ref{v00:t}) to~(\ref{ddv:o}) one gets the
gap equation
\begin{eqnarray} \label{gap:f}
\Omega^2 = \tilde{\omega}^2 + \frac{\lambda}{4 \Omega}
+\frac{\lambda F^2}{2\Omega^4},
\end{eqnarray}
which has only one positive root for $\Omega$. Note that
this gap equation exactly reproduces the result of
Ref.~\cite{spkim} if one turns off the external force
($F=0$).

The expectation values of operators with respect to the
ground states defined by~(\ref{0:I})are
\begin{eqnarray} \label{qp:vac}
\langle 0|\hat{q}(t)|0 \rangle &=& \frac{F}{\sqrt{m(t)}
    \Omega^2} \nonumber , \\
\langle 0|\hat{p}(t)|0 \rangle &=&  -\frac{\dot{m}(t) F}{2
    \Omega^2 \sqrt{m(t)} }   \nonumber , \\
\langle 0|\hat{q}^2(t)|0 \rangle &=& \frac{1}{2 m(t)
\Omega}
    +\frac{F^2}{m(t) \Omega^4}, \\
\langle 0|\hat{p}^2(t)|0 \rangle &=& \frac{m(t)\Omega}{2 }
   \left[1+\frac{\dot{m}^2(t)}{4 \Omega^2 m^2(t)}
   \right]
   +\frac{\dot{m}^2(t) F^2}{4 \Omega^4 m(t)},
   \nonumber \\
\langle 0|(\Delta \hat{q})^2(\Delta\hat{p})^2|0 \rangle &=&
\frac{1}{4}\left[1+ \frac{\dot{m}^2(t)}{4 \Omega^2 m^2(t)}
\right] . \nonumber
\end{eqnarray}
Note that the state $|0\rangle$ satisfies the minimal
uncertainty relation. In fact, $|0\rangle $ is the minimum
energy state of the Fock space. The force dependence on the
uncertainty enter only through the frequency indirectly.

Now let us go into the main problem in this section: a
time-dependent anharmonic oscillator with the driving force
linearly increasing to a maximum value $f$. For simplicity
of the problem, we remove the time-dependence of the
parameters  except for the force.

As a typical type of linear driving force, we consider the
force of the form,
\begin{eqnarray}\label{f:t}
F(t) = \left\{
\begin{tabular}{lll}
   0, &  $t < 0$ ,\\
   $\displaystyle \frac{f t}{\Delta}$, &
       $0 \leq t < \Delta$ , \\
   $f$, &  $t \geq \Delta$. \\
\end{tabular} \right.
\end{eqnarray}
For notational simplicity, we use $t_- ~(=0)$ and
$t_+~(=\Delta)$ when it is convenient. The $x(t)$ in
Eq.~(\ref{x:fv}) then becomes
\begin{eqnarray}\label{x:t}
x(t) = - \int^t_{-\infty} F(t') \tilde{v}(t') dt' = \left\{
\begin{tabular}{lll}
   0, &  $t < 0$ ,\\
   $\displaystyle -\frac{ft}{\Delta} \int_0^t
         \tilde{v}(t') dt' +
   \frac{f}{\Delta}\int^t_0 dt'
       \int^{t'}_0 dt'' \tilde{v}(t'')$, &
        $0 \leq t <\Delta$,\\
   $\displaystyle -f \int_0^t
         \tilde{v}(t') dt' +
   \frac{f}{\Delta}\int^{\Delta}_0 dt'
       \int^{t'}_0 dt'' \tilde{v}(t'')$,
    &  $t \geq \Delta$. \\
\end{tabular} \right.
\end{eqnarray}

Now let us solve Eq.~(\ref{ddv:o}) at each regions of time.
For $t<0$, the external force vanishes, and we find an
exact solution of Eq.~(\ref{ddv:o}) which describes the
evolution of the ground state of the oscillator:
\begin{eqnarray}\label{v_0}
\tilde{v}_-(t)= \frac{1}{\sqrt{2 \Omega_-}} e^{-i \Omega_-
t},
\end{eqnarray}
where $\Omega_-$ denotes the value of $\Omega(t)$ in
Eq.~(\ref{gap:f}) with no external force, $F=0$.

For $t>0$, we cannot find exact solution of the full
equation~Eq.~(\ref{ddv:o}). Instead, we try to find
approximate solution based on the WKB-type solution. $x(t)$
is proportional to $f$ and the $|v x^*-v^* x|^2$ term in
Eq.~(\ref{oeff:v}) would be $O(\lambda f^2/\omega^6 \cdot
\omega^2)$.  Therefore, we introduce the approximation for
$f^2 \lambda \ll \Omega^{6}$ and find the solution for
$\tilde{v}(t)$ up to linear order in $\lambda
f^2/\Omega^6$.
 Any types of forces can
be treated approximately if it satisfies $\lambda f^2 \ll
8\tilde{\omega}^6$. Note that since it restricts only the
product of the coupling and the driving force squared, the
driving force need not be small. The present method used in
this paper can cover any time-dependent external forces
which approaches to a finite value at $t \rightarrow
\infty$ if the coupling is small enough, which is usually
required even in  the absence of driving force. One may try
to find an approximate solution of Eq.~(\ref{ddv:o}) by
considering linearized equation for $\tilde{v}(t)-
\tilde{v}_-(t)$ and by using $\tilde{v}_-(t)$ as a zeroth
order solution. But this approach based on $\tilde{v}_-(t)$
fails to give appropriate description of the system because
the size of the perturbation linearly increases as
$\Omega\Delta$ at $t=\Delta$, which gives divergent result
in the adiabatic limit ($\Delta \rightarrow \infty$). In
fact, more appropriate zeroth order solution is the WKB
type:
\begin{eqnarray} \label{v0:t}
v_0(t) = \frac{1}{\sqrt{2\Omega(t)}}e^{-i \psi(t)},
\end{eqnarray}
where $\psi(t) = \int_0^t \Omega(t') dt'$, and the
frequency $\Omega(t)$ is determined by the gap
equation~(\ref{gap:f}), in which the driving force $F(t)$
is now time-dependent. Let us set the two values of
frequencies, $\Omega(0)$ and $\Omega(\Delta)$ as
$\Omega_\mp$, which are determined by
\begin{eqnarray} \label{Opm}
\Omega_-^2 = \tilde{\omega}^2 + \frac{\lambda}{4 \Omega_-}, ~~
\Omega_+^2 = \tilde{\omega}^2 + \frac{\lambda}{4 \Omega_+} +
\frac{\lambda f^2 }{2 \Omega_+^4} .
\end{eqnarray}
For later uses let us define two dimensionless constants:
\begin{eqnarray}\label{epsilon}
\nu= \frac{\lambda }{16 \Omega_-^3}, ~~\epsilon=
\frac{\lambda f^2}{8 \Omega_-^6(1+2\nu)} .
\end{eqnarray}
The phase in the intermediate time between $0 \leq t <
\Delta$ can be calculated up to $O(\epsilon)$ as:
\begin{eqnarray}\label{psi}
\psi(t) = \int_0^t\Omega(t')d t'= \Omega_- t
+\frac{2\epsilon \Omega_- \Delta}{3} \frac{t^3}{\Delta^3},
~ ~ 0 \leq t \leq \Delta.
\end{eqnarray}
Now we replace $\tilde{v}(t)$ in $\left|\tilde{v}(t)
x^*(t)-v^*(t) x(t)\right|^2$ term of Eq.~(\ref{oeff:v}) by
its zeroth order solution $v_0(t)$. Note that we do not
replace $|\tilde{v}(t)|^2$ term by its zeroth order
solution, since we now expand the solution in $\epsilon=
\lambda f^2/(8\Omega^6)$. Therefore $|\tilde{v}(t)|^2$ term
is zeroth order in this parameter. We calculate all
physical quantities up to this order. The equation of
motion~(\ref{ddv:o}) then becomes
\begin{eqnarray}\label{eom:g}
\ddot{\tilde{v}}(t) +\tilde{\omega}^2(t)\tilde{v}(t)
+\frac{\lambda}{2}\left\{|\tilde{v}(t)|^2
         +\frac{f^2}{\Omega^4(t)}\bar{f}^2(t)
          \right\}
\tilde{v}(t) = 0,
\end{eqnarray}
where $\bar{f}(t)$ is a dimensionless function defined by
\begin{eqnarray}\label{vx}
\bar{f}(t)&=&-i\frac{\Omega^2(t)}{f} [\tilde{v}(t)
x^*(t)-\tilde{v}^*(t) x(t)]  \nonumber \\
&\simeq &-i \frac{\Omega^2(t)}{f}[\tilde{v}(t)
x^*(t)-\tilde{v}^*(t) x(t)]_{\tilde{v}(t)=v_0(t)}  \\
&\simeq&\left\{\begin{tabular}{ll}
   0, &  $t < 0$ ,\\
   $\displaystyle \frac{1}{\Delta \Omega(t) }
     \left[ \psi(t)
     - \sin\psi(t)\right] ,$ &
        $0 \leq t <\Delta$,\\
   $\displaystyle
     1- \frac{\sin(\Omega_+\Delta/2)}{\Omega_+\Delta/2}
      \cos\Omega_+ \left( t - \frac{\Delta}{2}\right)
   ,$
    &  $t \geq \Delta$. \\
\end{tabular} \right.  \nonumber
\end{eqnarray}
During the calculation, one may find the following term,
\begin{eqnarray}\label{dddd}
-\frac{\sin\left[(\psi(\Delta)-\Omega_+
 \Delta)/2\right]}{\Delta\Omega/2} \cos\left[\Omega_+ t +
 \Omega_+ \Delta/2-\psi(\Delta)/2 \right],
\end{eqnarray}
added to the last line of Eq.~(\ref{vx}). This term can be
removed by the following arguments: First, consider the
case $\Delta \Omega_- \ll 1/\epsilon $. Then, the argument
of the sine function, $\displaystyle
\frac{\psi(\Delta)-\Omega_+\Delta}{2}=- \frac{1}{3}
\epsilon \Omega_-\Delta$ in Eq.~(\ref{dddd}) is
$O(\epsilon)$, which makes above terms to be $O(\epsilon)$.
On the other hand, if $\Delta \gg 1$, the denominator in
Eq.~(\ref{dddd}) diverges. Therefore, this term can be set
to zero safely to $O(\epsilon^0)$. There appears many
similar terms in the calculations, which we simply omit
without explicitly referring to it.

Because the driving term in Eq.~(\ref{eom:g}) is linear in
$\tilde{v}(t)$, the driving term in the potential which
gives Eq.~(\ref{eom:g}) is quadratic in $\tilde{v}(t)$. If
there is no higher power of $\tilde{v}(t)$ in the potential
one cannot prevent the solution resonating to the driving
source and becoming divergent, which is unwanted result.
Therefore, the higher power term in the potential is
necessarily needed, which is the $\lambda |\tilde{v}(t)|^2$
term in Eq.~(\ref{eom:g}). Even though we cannot solve this
non-linear time-dependent differential equation exactly, it
is possible to obtain the general properties of the
solution by separating $\tilde{v}(t)$ into its size $r(t)$
and phase $\phi(t)$:
\begin{eqnarray}\label{v:r,amp}
\tilde{v}(t) = r(t) e^{-i \phi(t)}.
\end{eqnarray}
Then the equation of motion~(\ref{eom:g}) is reduced to the
following two equations:
\begin{eqnarray}\label{ddr:0}
&&\ddot{r}(t)-\frac{L^2}{r^3(t)}+ \frac{\lambda}{2} r^3(t)
+ \left\{\tilde{\omega}^2+\frac{\lambda f^2}{2 \Omega^4}
 \bar{f}^2(t) \right\} r(t) =0, \\
&& r^2(t) \dot{\phi}(t) = L^2, \label{rphi:L}
\end{eqnarray}
where $L$ is a constant of motion determined by
\begin{eqnarray}\label{L,r0}
L&=& r^2(t) \dot{\phi}(t)=r^2(\Delta) \dot{\phi}(\Delta)
=r^2(0)\dot{\phi}(0)=\frac{1}{2}.
\end{eqnarray}
Note that Eq.~(\ref{L,r0}) is very similar to the
conservation law of the angular momentum. Moreover,
Eq.~(\ref{ddr:0}) is very similar to the radial equation of
2-dimensional rotating particle, except that the $r$
dependence on Eq.~(\ref{ddr:0}) is $1/r^3$ rather than
$1/r^2$. The effective potential, which gives
Eq.~(\ref{ddr:0}), can be written as
\begin{eqnarray}\label{V:r}
V(r,t) &=& \frac{L^2}{2 r^2}+ \frac{\lambda}{8}r^4 +
\left\{\tilde{\omega}^2+\frac{\lambda f^2}{2 \Omega^4}
 \bar{f}^2(t) \right\}
  \frac{r^2}{2}   \\
  &=& V_{\pm}(r) +\frac{\lambda f^2}{2 \Omega^4}
 \left[\bar{f}^2(t)- f_{\pm}
  \right]
  \frac{r^2}{2}   , \nonumber
\end{eqnarray}
where $f_+ = 1$, $f_-=0$, and $V_\mp(r)=\frac{L^2}{2 r^2}+
\frac{\lambda}{8}r^4 + \left[\tilde{\omega}^2+\frac{\lambda
f^2}{2 \Omega^4}
 f_\pm \right]\frac{r^2}{2} $ are the adiabatic
 ($\Delta \rightarrow \infty$) potential at
$t < 0$, and $t>\Delta$ respectively. If the system is
quasi-static ($\Omega \Delta \gg 1$), the function
$\bar{f}(t)$ in the potential~(\ref{V:r}) monotonically
increases from $0$ ($t=0$) to $1$ ($t= \Delta$) in time,
which define $V_\pm(r)$. The two stable circular orbit
$r=r_-$ (at $t<0$) and $r_+$ (at $t>\Delta$) are determined
by the formula $V'_\pm(r)=0$, where prime denotes the
derivative with respect to $r$, and are given by
\begin{eqnarray}
r_-= \frac{1}{\sqrt{2 \Omega_-}}, ~~ r_+= \frac{1}{\sqrt{2
\Omega_+} } .
\end{eqnarray}
The evolution of $(r(t),\phi(t))$, therefore, can be seen
as an orbit change of a circulating particle by a varying
frequency with somewhat odd angular momentum dependence.

Now let us approximate the potential up to quadratic order
in $\epsilon$ near $r=r_\pm$:
\begin{eqnarray}\label{V:r2}
V(r,t) \simeq V_{eff}(r,t) &=& V_{\pm}(r_{\pm}) +
\frac{1}{2}V''_\pm(r_{\pm})(r-r_{\pm})^2 -\frac{\lambda f^2
r_{\pm}}{2 \Omega^4(t)} J(t)(r-r_{\pm}) , \nonumber
\end{eqnarray}
where we have written the time-dependent term up to the
first order in $r-r_\pm$ since it is of $O(\epsilon)$
already and the source
\begin{eqnarray}\label{J}
J(t) = \theta(t)[\theta(\Delta-t) - \bar{f}^2(t)],
\end{eqnarray}
is dimensionless.

Let the orbit difference be
\begin{eqnarray}\label{rho}
 \rho_{I-}(t)&=& \frac{1}{\epsilon_- r_-} [r(t) -
      r_{-}], ~~~ 0\leq t<\Delta , \\
 \rho_{I+}(t)&=& \frac{1}{
     \epsilon_+ r_+} [r(t) -
r_{+}], ~~~  t \geq \Delta ,  \nonumber
\end{eqnarray}
where $\displaystyle \epsilon_\pm = \frac{\lambda f^2}{2
\Omega_\pm^4\bar{\omega}^2_\pm}$ is a dimensionless
expansion parameter, and $\bar{\omega}^2_\pm=
V''_\pm(r_{\pm})$. The equation of motion for $\rho(t)$ is
given by
\begin{eqnarray}\label{eom:rho}
\ddot{\rho}_{I\pm}(t) + \bar{\omega}^2_\pm \rho_{I\pm}(t)=
 \bar{\omega}^2_\pm J(t),
\end{eqnarray}
where the subscripts $\pm$ are used for $t \geq \Delta$ and
$0\leq t<\Delta$, respectively. This is just the equation
of motion of a harmonic oscillator with a driving force,
and its inhomogeneous solution can be written explicitly
using the Green's function method:
\begin{eqnarray}\label{rhoI}
\rho_{I\pm}(t)&=&\int_{t_\pm}^t \bar{\omega}_\pm dt' J(t') \sin
\bar{\omega}_\pm(t-t').
\end{eqnarray}
Note that this inhomogeneous solution satisfies
$\rho_{I\pm}(t_\pm)=0=\dot{\rho}_{I\pm}(t_\pm)$. We present
the explicit form of $\rho_{I\pm}(t)$ in Appendix A.

With this $\rho_{I\pm}(t)$ in hand, it is easy to write the
solution for $r(t)$ which satisfies the appropriate
boundary condition at $t=t_\pm$:
\begin{eqnarray}\label{sol:r}
r(t)&=& r_{\pm}+ \epsilon r_{\pm} \rho_{I\pm}(t) +
[r(t_\pm)-r_{\pm}]\cos \bar{\omega}_\pm(t-t_\pm) +
\frac{\dot{r}(t_\pm)}{\bar{\omega}_\pm } \sin
\bar{\omega}_\pm(t-t_\pm) \\
&=& r_{\pm}+s_\pm(t)
 ,\nonumber
\end{eqnarray}
where $s_\pm(t)$ denotes the time-dependent parts of $r(t)$
in the region $t \geq \Delta $ and $0 \leq t <\Delta$,
respectively. We have omitted the sign $\pm$ in $\epsilon$
here and later since $\epsilon_\pm$ are the same to this
order. Usually, the homogeneous terms are first order in
the expansion coefficient ($\epsilon$ here).

The boundary condition at $t=0$ is
\begin{eqnarray}\label{bc:0}
r(0)= \frac{1}{\sqrt{2\Omega_-}}, ~~\dot{r}(0)=0,~~
\phi(0)=0, ~~ \dot{\phi}(0)=\Omega_- ,
\end{eqnarray}
where $r(0)$ was determined by Eq.~(\ref{comm:v}). After
matching these boundary conditions we get
\begin{eqnarray}\label{L}
L=\frac{1}{2}, ~~~r_-=r(0), ~~~\bar{\omega}_-^2= 4
\tilde{\omega}^2+ \frac{3 \lambda}{2 \Omega_-} =4
\Omega_-^2(1+2\nu),
\end{eqnarray}
and
\begin{eqnarray}\label{r:mid}
r(t) = \frac{1}{\sqrt{2 \Omega_-}}\left[1+
\epsilon\rho_{I-}(t)\right].
\end{eqnarray}

The solution for $r(t)$ in the region $t\geq \Delta$, which
matches the boundary condition at $t = \Delta$, is given by
\begin{eqnarray}\label{r:g}
r(t)&=&  r_+ + s_+(t)  \\
 &=& r_+ +  \epsilon r_+ \rho_{I+}(t) +
\frac{\epsilon}{\sqrt{2\Omega_-}}
\left[\rho_{I-}(\Delta)+1\right]\cos
\bar{\omega}_+(t-\Delta) \nonumber \\
& & + \frac{\epsilon}{\sqrt{2\Omega_-}}
\frac{\dot{\rho}_{I-}(\Delta)}{\bar{\omega}_+ } \sin
\bar{\omega}_+(t-\Delta) ,\nonumber
\end{eqnarray}
where $\displaystyle \bar{\omega}_+^2= V_+''(r_+)=4
\Omega_+^2(1+ 2\nu) $.

In the region $0\leq t < \Delta $, the phase function can
be written (See Appendix B for details.) as
\begin{eqnarray} \label{phi:psi}
\phi(t)&=& \psi(t) + \epsilon \delta \phi_-(t).
\end{eqnarray}
where $\psi(t)$ is given by~(\ref{psi}) and $\delta
\phi_-(t)$ is a finite oscillating function defined in
Eq.~(\ref{phi:psi}). With these results we obtain
$\tilde{v}(t)$ in the region, $0 \leq t <\Delta$,
\begin{eqnarray}\label{v:mid}
\tilde{v}(t) = v_-(t) + \epsilon v_-(t) w_-(t),
\end{eqnarray}
where
\begin{eqnarray}\label{v_-}
v_-(t) &=& \frac{1}{\sqrt{2\Omega(t)}} e^{-i \psi(t)},
\end{eqnarray}
and $w(t)$ is a function of $O(1)$. Then $x(t)$ in
Eq.~(\ref{x:fv}) is given by
\begin{eqnarray}\label{x:mid}
x(t) =- \frac{f v_-(t)}{\Delta \Omega_-^2}\left[1 + i
\Omega_- t- e^{i \psi(t)}\right]\left(1+ \epsilon
\frac{t^2}{\Delta^2}\right)
 + \frac{f v_-(t)}{\Delta} \epsilon x_1(t)
 ,
\end{eqnarray}
where $x_1(t)$ is defined in Eq.~(\ref{x_1}) in Appendix B.
Now we compute the expectation value of operators:
\begin{eqnarray}
\langle 0|\hat{q}(t)|0\rangle &=& \frac{f}{\Delta
\Omega^3}\left\{ [\Omega t - \sin \psi(t)]
 \left\{ 1+ \epsilon \Re[w(t)] \right\}- \epsilon
 \left[1-\cos\psi(t) \right] \Im[w(t)]-\epsilon
  \Im(x_1)\right\}
  ,\nonumber \\
\frac{\langle 0|\hat{p}(t)|0\rangle}{m} &=& \frac{f}{\Delta
\Omega^2} \left\{ [1-\cos\psi(t)] \left[1+ \frac{2\epsilon
t^2 }{\Delta^2} + \epsilon \Re\left( w(t) + i
\frac{\dot{w}(t)}{\Omega}\right) \right] \right. \\
&-& \left. \epsilon \Re[x_1(t)] + \epsilon [\Omega t - \sin
\psi(t)] \Im\left(w(t)+ i \frac{\dot{w}(t)}{\Omega}\right)
\right\}
 .  \nonumber
\end{eqnarray}

In the region $t> \Delta$ we get quite good approximation
of the phase up to $O(\epsilon)$:
\begin{eqnarray}\label{phi:approx}
\phi(t) &\simeq& \frac{t-t_1}{2\bar{r}^2} +
 \epsilon \delta \phi_+(t),
\end{eqnarray}
where $\delta \phi(t)$ is an oscillating function of
$O(1)$, and $\bar{r}=r_{+}\left[1+ \epsilon
\frac{\sin^2\Omega \Delta/2}{(\Omega \Delta/2)^2}\right]$
is given in Eq.~(\ref{r:avg}). Then $\tilde{v}(t)$ and
$x(t)$ can be written to this order,
\begin{eqnarray}\label{v:t}
\tilde{v}(t) &\simeq & \left[1 + \epsilon w_+(t) \right]
v_+(t) =\left[1 +\epsilon w_+(t) \right] \bar{r}
e^{-i(t-t_1)/(2\bar{r}^2)} , \\
x(t) &\simeq & -2i \bar{r}^2 f v_+(t) \left[1- \frac{\sin
[\Delta/(4\bar{r}^2)]}{\Delta/(4\bar{r}^2)} e^{i
\frac{2t-\Delta}{4 \bar{r}^2}}- \epsilon y
e^{i\frac{t-\Delta}{2\bar{r}^2}} + \epsilon z(t) \right] ,
\end{eqnarray}
where $t_1$, $w_+(t)$, $y$, and $z(t)$ are given in
Appendix B. Using these we calculate the expectation values
of operators $\hat{p}(t)$ and $\hat{q}(t)$ at $t> \Delta$
which shows the effects of the driving force. The
expectation value of $q(t)$ and $p(t)$ in this order are
given by
\begin{eqnarray}
&&\langle 0|\hat{q}(t)|0\rangle    \label{q:0} \\
&& ~= 4 f \bar{r}^4 \left\{1 - \frac{\sin \Delta/(4
\bar{r}^2)}{\Delta/(4 \bar{r}^2)} \cos
\frac{t-\Delta/2}{2\bar{r}^2} -\epsilon
\left[e^{-i\frac{t-\Delta}{2 \bar{r}^2}}  y^*+
e^{i\frac{t-\Delta}{2 \bar{r}^2}}y +
-z^*(t)-z(t)\right]\right\}
, \nonumber \\
            \label{p:0}
&&\frac{\langle 0|\hat{p}(t)|0\rangle }{m} = 2 f
   \bar{r}^2
\left\{ \left[ 1+
    \epsilon \Re\left(w_+(t)+2i\bar{r}^2 \dot{w}_+(t)\right)
    \right]\frac{\sin \Delta/(4
   \bar{r}^2)}{\Delta/(4 \bar{r}^2)} \sin
   \frac{t-\Delta/2}{2\bar{r}^2} \nonumber \right. \\
&& ~~~~\left. + \epsilon \Im\left(w_+(t)+2i\bar{r}^2
\dot{w}_+(t)\right) \left(1-\frac{\sin \Delta/(4
   \bar{r}^2)}{\Delta/(4 \bar{r}^2)} \cos
   \frac{t-\Delta/2}{2\bar{r}^2} \right) \right\} \\
&&~~~~ +2 \epsilon f \bar{r}^2 \left[ ye^{i
   \frac{t-\Delta}{2 \bar{r}^2}}-y^* e^{-i
   \frac{t-\Delta}{2 \bar{r}^2}}- z(t)+z^*(t)
 \right] .  \nonumber
\end{eqnarray}

The fluctuations of $\hat{q}(t)$ and $\hat{p}(t)$ are given
by
\begin{eqnarray} \label{q2p2:vac}
\langle 0|[\Delta \hat{q}(t)]^2|0\rangle &=&
|\tilde{v}(t)|^2
  = \frac{\bar{r}^2}{m} \left[1+ 2 \epsilon g(t) \right]  , \\
\langle 0|[\Delta\hat{p}(t)]^2|0\rangle &=&
m^2|\dot{\tilde{v}}(t)|^2
  =\frac{m}{4 \bar{r}^2} [1-2 \epsilon g(t) ],
\nonumber
\end{eqnarray}
where $g(t)$ is a purely oscillating function of time of
order $O(1)$ defined in Appendix B~(\ref{s:r}). Therefore,
the uncertainty to order $\epsilon$ becomes
\begin{eqnarray}\label{uncertainty}
\langle (\Delta \hat{p})^2(\Delta \hat{q})^2 \rangle =
\frac{1}{4} + O(\epsilon^2).
\end{eqnarray}
If one compares Eq.~(\ref{q2p2:vac}) with
Eq.~(\ref{qp:vac}) one can naturally identify that
$1/(2\bar{r}^2)$ plays the role of frequency of the
oscillator.

The main effects of the driving force to the dispersions of
$\hat{q}(t)$ and $\hat{p}(t)$ come in two fold: First,
driving force makes the dispersions fluctuating $[s(t)]$.
Secondly, it permanently squeezes the wave-packet by
distorting its frequency, $[1/(2\bar{r}^2)]$. Note that the
adiabatic limit, $\bar{r}=r_+$, is the least squeezing
state of the system. An interesting point is that there are
other systems with $\Delta/(4\bar{r}^2) =n \pi$ which
reveals the same least squeezing as the adiabatic limit.
The interpretation on this phenomenon is given in the last
section.

\section{ Weak coupling limit }

In this section we consider the weak coupling limit ($\nu
=\lambda/(16\Omega^3) \ll 1$) of the time-dependent
anharmonic oscillator with driving force given
by~Eq.~(\ref{f:t}). One should note that there are terms of
$O(1/\nu )$ given in $\rho_{I\pm}(t)$ in the Appendix A, if
the coupling is very small. The inhomogeneous solutions are
approximated in the weak coupling limit as
\begin{eqnarray}\label{rho:weak}
\rho_{I-}(t) &=&- \frac{1}{4\nu\Delta^2 \Omega_-^2}
\left[\cos \bar{\omega}_-t -\cos 2 \psi(t) \right]
-\frac{\psi^2(t)}{\Delta^2 \Omega_-^2} + g'(t), \\
\rho_{I+}(t) &=& -\frac{\sin^2(\Omega_+\Delta /2)}{2 \nu
(\Omega_+ \Delta /2)^2} \sin \frac{(\bar{\omega}_++2
\Omega_+)t- \bar{\omega}_+ \Delta }{2}
\sin\frac{(\bar{\omega}_+ -2 \Omega_+)(t-\Delta) }{2}  \\
&& -\frac{\sin^2(\Omega_+\Delta /2)}{2 (\Omega_+ \Delta
/2)^2} + g''(t) , \nonumber
\end{eqnarray}
where $g'(t)$ and $g''(t)$ are functions, whose integrals
over time from $0$ to $t$ are of $O(\epsilon^0)$. We ignore
$O(\lambda)$ terms in $\tilde{v}(t)$ and discuss these
terms later. To this order, $\tilde{v}(t)$ is given by
\begin{eqnarray}
\tilde{v}(t) =\left \{ \begin{tabular}{ll}
  $ \tilde{v}_-(t)
     $, &
       $ 0 \leq t < \Delta , $\\
 $\tilde{v}_+(t)$,   & $ t \geq \Delta , $\\
\end{tabular} \right.
\end{eqnarray}
where
\begin{eqnarray}\label{v:weak}
\tilde{v}_-(t)&=& v_-(t)\left\{1 + i \beta
  \sin\left[\psi(t)-\frac{\bar{\omega}_-t}{2}\right]
   e^{i [\psi(t)+\bar{\omega}_-t/2]} \right\}, \\
\tilde{v}_+(t)&=&\left[1+ \frac{i
        \dot{s}_-(\Delta)}{\bar{r}\bar{\omega}_- }
         \right]v_+(t)
     + \left[s_-(\Delta)- \frac{i
     \dot{s}_-(\Delta)}{\bar{\omega}_+}
     \right]e^{i(\bar{\omega}_+ -
     \Delta)t}\frac{v_+(t)}{\bar{r}} \\
&+& 2i \alpha v_+(t) e^{i\psi_+(t)} \sin\psi_-(t)
       , \nonumber
\end{eqnarray}
where the functions $\psi_\pm(t)$, $v_+(t)$, and the
coefficients are defined by
\begin{eqnarray}\label{alpha}
\psi_+(t)&=&\left(\frac{\bar{\omega}_+}{2}+
\Omega_+\right)(t-\Delta)+ \Omega_+\Delta ,~~~
\psi_-(t)=\left(\frac{\bar{\omega}_+}{2}-
\Omega_+\right)(t-\Delta), \\
 v_+(t) &=& \bar{r}
e^{-i\bar{\Omega}(t-t_1) }, ~~ \alpha
=\frac{f^2\sin^2(\Omega_+ \Delta/2)}{2\Omega_+^3( \Omega_+
\Delta/2)^2},~~ \beta= \frac{f^2}{\Delta^2 \Omega_-^5},~~
\bar{\Omega}=\frac{1}{2\bar{r}^2}.  \nonumber
\end{eqnarray}
The initial condition~(\ref{r:g}) at $t=\Delta$ is given by
\begin{eqnarray}\label{s:Delta}
s_-(\Delta) &=& -r_-\beta \left\{\sin
\left[\frac{\bar{\omega}_-\Delta}{2}+ \psi(\Delta)\right]
\sin\left[ \frac{\bar{\omega}_-\Delta}{2}-
\psi(\Delta)\right]\right\},  \\
\dot{s}_-(\Delta) &=& - \beta \Omega_- r_-\left\{\cos
\left[\frac{\bar{\omega}_-\Delta}{2}+ \psi(\Delta)\right]
\sin\left[ \frac{\bar{\omega}_-\Delta}{2}-
\psi(\Delta)\right]\right\}.
\end{eqnarray}
Note that these terms can be removed by the following
arguments: First, consider the case $\Omega\Delta \ll 1/\nu
$. Then, the argument of the sine function,
$\frac{\bar{\omega}_-\Delta}{2}- \psi(\Delta)=\Omega_-
\Delta\left(\nu- \frac{2\epsilon}{3}\right)$ in
Eq.~(\ref{s:Delta}) is $O(\epsilon)$ which makes above
terms to be $O(\epsilon)$. On the other hand, if $\Delta
\gg 1$, $\beta$ goes to zero in this case. Therefore, the
homogeneous terms in the region $t> \Delta$ can be set to
zero safely to this order. There appears many similar terms
in the calculations, which we simply omit without
explicitly referring to it.

 What we want to know is the expectation values of
operators at $t \geq \Delta$. Therefore we should calculate
$x(t)$ in this region of time. At $t=\Delta$, by
integrating $-F(t)\tilde{v}_-(t)$ in Eq.~(\ref{v:weak}), we
have
\begin{eqnarray}\label{x:Delta}
x(\Delta) &=& -\frac{if v_-(\Delta)}{\Omega_-} \left[1-
\frac{\sin \psi(\Delta)/2}{\Omega_- \Delta /2} e^{i \psi(
\Delta)/2} \right].
\end{eqnarray}
In general, for $t \geq \Delta$,
\begin{eqnarray}\label{x:t:weak}
x(t) = x(\Delta) -f \int_\Delta^t \tilde{v}(t') dt'= x_+(t)
+ x(\Delta)-x_+(\Delta),
\end{eqnarray}
where
\begin{eqnarray}\label{fv}
x_+(t) &=& -\int_0^t F(t) \tilde{v}_+(t') dt' \\
&=&  -\frac{if v_+(t)}{\Omega_+} \left\{1-
\frac{\sin(\Omega_+ \Delta/2)}{\Omega_+ \Delta/2} e^{i
\bar{\Omega}(t-\Delta/2)}
\right. \nonumber \\
& -& \left. 2i\alpha \sin\psi_-(t) e^{i\psi_+(t) }
 \right\} . \nonumber
\end{eqnarray}

At $t=\Delta$ the following equality holds:
\begin{eqnarray}\label{v+:v-}
\tilde{v}_-(\Delta)=v_-(\Delta)= v_+(\Delta)+O(\epsilon).
\end{eqnarray}
We therefore have,
\begin{eqnarray}\label{x:t}
x(t) = x_+(t) + O(\epsilon).
\end{eqnarray}
The expectation value of $\hat{q}(t)$ is then given by
\begin{eqnarray}\label{pq:weak}
\langle 0|\hat{q}(t)|0\rangle &=&\frac{f}{\sqrt{m}
\Omega_+^2} \left[1+ 2 \alpha \cos \psi_+(t)
\sin\psi_-(t) \right]  \nonumber \\
 && \cdot \left[ 1 -\frac{\sin(\Omega_+
\Delta/2)}{\Omega_+ \Delta /2} \cos
[\bar{\Omega}(t-\Delta/2)]+ 2 \alpha \cos \psi_+(t) \sin
\psi_-(t) \right]  \nonumber
 \\
&& -\frac{2\alpha f}{\sqrt{m} \Omega_+^2}
\sin\psi_+(t)\sin\psi_-(t) \left[
 \frac{\sin(\Omega_+\Delta/2)}{\Omega_+\Delta/2}
 \sin[\bar{\Omega}(t-\Delta/2)]  \right. \\
 &&\left.  - 2 \alpha \sin\psi_+(t)
 \sin\psi_-(t) \right]
+ O(\epsilon). \nonumber
\end{eqnarray}
The solution is composed of the multiplication of two
parts: the rapidly oscillating part, $\psi_+(t)$, and
slowly oscillating part, $\psi_-(t)\simeq \nu(t-\Delta)$.
The Eq.~(\ref{pq:weak}) is exact answer to all order in
$\alpha$ for $\alpha \gg \epsilon$. The correction to
Eq.~(\ref{pq:weak}) begins only at the $O(\epsilon)$. It is
not needed to calculate the expectation value of
$\hat{p}(t)$, because it is simply given by $\displaystyle
\langle \hat{p}(t)\rangle= m \frac{d \langle
\hat{q}(t)\rangle}{d t} $. This proves that the adiabatic
theorem holds in the absence of $O(\epsilon)$ correction.
The $O(\epsilon) $ corrections are given by Eq.~(\ref{q:0})
with the replacement $g(t)$ in Eq.~(\ref{r2}) by $g'(t)$,
$g(t)$ in Eq.~(\ref{r2:large}) by $g''(t)$. The adiabatic
limit of weak coupling expansion is written explicitly in
Eq.~(\ref{adiabatic:vac}). The dispersions are given by
\begin{eqnarray} \label{q2p2:weak}
\langle 0|[\Delta \hat{q}(t)]^2|0\rangle &=&
  \frac{1}{2m\bar{\Omega}} \left[ 1- 4 \alpha \sin\psi_+(t)
   \sin\psi_-(t) + 4 \alpha^2 \sin^2 \psi_-^2(t)\right]
    , \\
\langle 0|[\Delta \hat{p}(t)]^2|0\rangle &=&
   \frac{m\bar{\Omega}}{2} \left[ 1+ 4 \alpha \sin\psi_+(t)
   \sin\psi_-(t) + 4 \alpha^2 \sin^2 \psi_-^2(t)\right]
    . \nonumber
\end{eqnarray}
Note that the uncertainty is $1/4$ to $O(\alpha)$ even
though the dispersions are fluctuating in $O(\alpha)$. In
the adiabatic limit, the uncertainty becomes exactly 1/4,
which is the minimal one.

The $O(\epsilon)$ calculation is not important except for
the adiabatic limit and the case $\Delta \Omega_+=n \pi$.
In the adiabatic case we explicitly calculate each
functions again up to $O(\epsilon)$ and we get, in the
region $t \geq \Delta$,
\begin{eqnarray}
\tilde{v}(t) &=& \frac{1}{\sqrt{2 \Omega_+}}
e^{-i\Omega_+(t-
\Delta)-i \psi(\Delta)}, \\
x(t)&=& \frac{-if \tilde{v}(t)}{\Omega_+} \left[1+4
\epsilon e^{-i \Omega_+(t-\Delta)}\right].
\end{eqnarray}
Therefore the expectation values of operators are given by
\begin{eqnarray} \label{adiabatic:vac}
\langle 0|\hat{q}(t)|0\rangle &=& \frac{f}{\Omega_+^2}
\left\{1+ 4 \epsilon \cos
[\Omega_+(t-\Delta)]\right\}, \\
\langle 0|\hat{p}(t)|0\rangle &=&-\frac{4 \epsilon
f}{\Omega_+^2} \sin [\Omega_+(t-\Delta)].
\end{eqnarray}
The dispersions are
\begin{eqnarray}\label{delQ:adiaWeak}
\langle 0|[\Delta \hat{q}(t)]^2|0\rangle &=& \frac{1}{2 m
\Omega_+}, \\
\langle 0|[\Delta \hat{p}(t)]^2|0\rangle &=& \frac{m
\Omega_+}{2} .
\end{eqnarray}
The uncertainty takes the minimal value. Therefore, in the
adiabatic limit of weak coupling expansion, the initial
ground state becomes a gaussian wave-packet, which is very
close to the new ground state except that the center of the
packet is oscillating, whose amplitude is of $O(\epsilon)$.

\section{Discussions}

We have formulated the quantization procedure of a
time-dependent driven anharmonic oscillator using the
Liouville-von Neumann approach, and then applied it to the
case of anharmonic oscillator with linearly varying driving
force. We have found a new gap equation for the driven
oscillator and examined whether or not the adiabatic
conjecture hold for the anharmonic oscillator. We also have
calculated the weak coupling limit of the system. It was
shown that the adiabatic conjecture breaks down at the
first order of the coupling and the gaussian ground state
evolves into a coherent state with the $O(\epsilon)$
correction of squeezing parameter.

The weak coupling expansion is most interesting because
there appears a new expansion parameter $\alpha \gg
\epsilon$~(\ref{alpha}), which are closely related with the
force gradient $f/\Delta$. In the adiabatic limit of weak
coupling expansion, all dynamics related to the new
parameter $\alpha$ disappears and only the very small
$O(\epsilon)$ correction remains. The expectation values of
operators in the adiabatic limit of the weak coupling, at
$t>\Delta$ are given in Eq.~(\ref{adiabatic:vac}). The
dispersion~(\ref{delQ:adiaWeak}) implies that the
wave-packet has the shape of a ground state of an
oscillator with frequency $\Omega_+$. This state of the
adiabatic limit is the coherent state with the coherence
parameter of $O(\epsilon)$. Explicitly, the invariant
operator can be rewritten as
\begin{eqnarray}\label{A:large}
\hat{A}&=& i \tilde{v}^*(t)\left\{\hat{p}(t)-i m
\Omega_+\left[ \hat{q}(t)- \frac{f}{m \Omega_+^2}\right]
\right\} +\frac{4 \epsilon f}{\Omega_+} e^{i \psi(\Delta)}
 \\
 &=& e^{i\Omega_+(t-
\Delta)+i \psi(\Delta)} \hat{B}(t) + \frac{4 \epsilon
f}{\Omega_+} e^{i \psi(\Delta)}, \nonumber
\end{eqnarray}
where $\hat{B}(t)$ is an annihilation operator of the
harmonic oscillator whose center of the potential is fixed
at $\frac{f}{\sqrt{m}\Omega_+^2}$ with natural frequency
$\Omega_+$. This observation naturally shows that all
initial states of the anharmonic oscillators are mapped
into the corresponding states of the new vacuum except for
$O(\epsilon)$ displacement.

There are infinitely many ways which gives states
satisfying Eq.~(\ref{adiabatic:vac}). If
$\Delta/(4\bar{r}^2)  = n \pi$ in
Eqs.~(\ref{q:0},\ref{p:0}), Eq.~(\ref{adiabatic:vac})
naturally holds and the corresponding final state is very
close to the new ground state. These arguments hold also in
the weak coupling limit of the
system~Eq.~(\ref{q2p2:weak}). The natural interpretation of
this phenomena is the following: After turning on the
driving force, the center of gaussian wave-packet starts to
move. It periodically passes the bottom of the potential,
and accidentally, if the driving force stop to increase
when the center stays the bottom of the potential, the
packet stop to move thereafter.

\vspace{0.5cm}
~\\
{\Large {\bf Acknowledgments}} \\
~\\
This work was supported in part by Korea Research
Foundation under Project number 99-005-D00010 (H.-C.K. and
J.H.Y.).

\vspace{0.5cm}
~\\
\newpage

{\Large \bf Appendix A. Source integration.} \vspace{0.5cm}

The Green's function in Eq.~(\ref{rhoI}) can be explicitly
integrated to
\begin{eqnarray}\label{rhoI-}
\rho_{I-}(t)&=& \int_0^t
dt'\frac{\bar{\omega}_-}{\Delta^2\Omega^2(t)}\left[\psi(t')
     - \sin\psi(t')\right]^2\sin\bar{\omega}_-(t-t') \\
&=&\frac{1}{2 \Delta^2 {\Omega}_-^2}
 \left\{-\left(1-4 \xi^2 \right)
      (1-\cos\bar{\omega}_-t) +
      \frac{2}{1-4\xi^2 }
      \sin\left[
       \frac{\bar{\omega}_-t}{2}+\psi(t) \right]
      ~\sin\left[
      \frac{\bar{\omega}_-t}{2}- \psi(t) \right]
      \right. \nonumber \\
& -&  \frac{16 \xi^2}{(1-\xi)^2(1+\xi)^2}
 \sin\left[
       \frac{\bar{\omega}_-t+\psi(t)}{2} \right]
      ~\sin\left[
      \frac{\bar{\omega}_-t- \psi(t)}{2} \right] \nonumber \\
 &+&\left. \frac{4}{1-\xi^2}\psi(t) \sin \psi(t)- 2\psi^2(t)   \nonumber
 \right\}  ,
\end{eqnarray}
where $\xi = \Omega_-/\bar{\omega}_-$. Note that this
inhomogeneous solution have a quadratically increasing term
in time. There is a term of $O(1/\lambda)$ in the $\lambda
\rightarrow 0$ limit because $(1-4 \xi^2) \simeq 2\nu$ in
this limit.

After integration of the Green's function $\rho_{I+}(t)$ is
given by
\begin{eqnarray}\label{rhoI+}
\rho_{I+}(t)&=&\int_\Delta^t dt'\bar{\omega}_+ J(t')
 \sin \bar{\omega}(t-t')  \\
 &=&
 \frac{2\sin \Omega_+ \Delta/2}{\Omega_+ \Delta/2}
\left[\frac{\sin\frac{(\bar{\omega}+\Omega_+)t-
\bar{\omega}\Delta }{2}
\sin\frac{(\bar{\omega}-\Omega_+)(t-\Delta)}{2} }{1-\chi} +
\frac{\sin\frac{(\bar{\omega}+\Omega_+)(t-\Delta)}{2}
\sin\frac{(\bar{\omega}-\Omega_+)t+\bar{\omega}\Delta}{2}
}{1+\chi}  \right] \nonumber \\
&-& \frac{\sin^2 (\Omega_+ \Delta/2)}{2(\Omega_+
\Delta/2)^2}
 \left[1- \cos \tilde{\omega}(t-\Delta)
 +\frac{\sin\frac{(\bar{\omega}+2\Omega_+)t-
     \bar{\omega} \Delta}{2}
     \sin\frac{(\bar{\omega}-2\Omega_+)(t
        -\Delta)}{2} }{1-2\chi} \right. \nonumber \\
&+& \left.
\frac{\sin\frac{(\bar{\omega}+2\Omega_+)(t-\Delta)}{2}
  \sin\frac{(\bar{\omega}-2\Omega_+)t+\bar{\omega}\Delta}{2}
   }{1+2\chi}  \right] ,  \nonumber
\end{eqnarray}
where $\chi= \Omega_+/\bar{\omega}_+$. The
solution~(\ref{rhoI+}) vanishes if the adiabatic limit,
$\Delta \Omega \rightarrow \infty$, is taken. It also
vanishes if the time interval of the driving force
satisfies $\Delta = n \pi/\Omega$. Note that there is a
term of $O(1/\lambda)$, if the coupling strength vanishes,
since $(1-2 \chi) \simeq \nu$.

\vspace{1.5cm}

{\Large \bf Appendix B. Calculation of $\tilde{v}(t)$ and
$x(t)$.} \vspace{0.5cm}

The phase factor $\phi(t)$ is given by
\begin{eqnarray}\label{phi:t}
\phi(t) = \int_0^t\frac{dt'}{2r^2(t')} .
\end{eqnarray}
Because we usually exponentiate this phase with $i$, one
should be cautious in approximating this integral by
separating the indefinitely growing part from the small
fluctuations. We calculate this integral for each regions
of time separately. To have a good approximation of the
integral in~(\ref{phi:t}) one should separate $r(t)$ by its
frequency and fluctuating parts. The integration of the
frequency part continuously increases in time, which gives
the phase of $\tilde{v}(t)$, and the integration of
fluctuating part should converge to a small value of
$O(\epsilon)$ so that its exponential can be Taylor
expanded.

In the region $0 \leq t < \Delta $, following the above
prescription with the explicit form of $\rho_{I-}(t) $
of~(\ref{rhoI-}), $r(t)$ can be written as
\begin{eqnarray}\label{r2}
r(t)= \frac{1}{\sqrt{2 \Omega_-}} \left[\left(1-
   \frac{\epsilon \psi^2(t)}{\Delta^2\Omega^2_-}
    \right) + \epsilon g(t) \right]
    = \frac{1}{\sqrt{2 \Omega(t)}} \left[1
    + \epsilon g(t) \right],
\end{eqnarray}
where $\psi(t)$ is given in Eq.~(\ref{psi}) and the
integral of $\displaystyle g(t)= \rho_{I-}(t) +
\frac{\psi^2(t)}{\Delta^2\Omega_-^2}$ in Eq.~(\ref{r2})
converges to a finite number for any value of $t$ or
$\Delta$. Expanding the right hand side of~(\ref{r2}) in
$\epsilon$ the phase integral~(\ref{phi:t}) is approximated
as
\begin{eqnarray}\label{phi:psi}
\phi(t)
 &\simeq & \psi(t) - 2 \epsilon \Omega_-
  \int _0^t g(t') dt'. \nonumber
\end{eqnarray}
With this calculation we obtain $\tilde{v}(t)$ in the
region $0 \leq t <\Delta$:
\begin{eqnarray}\label{v:mid}
\tilde{v}(t) = v_-(t) + \epsilon v_-(t) w(t),
\end{eqnarray}
where
\begin{eqnarray}\label{v_-}
v_-(t) &=& \frac{1}{\sqrt{2\Omega_-}} e^{-i \psi(t)}, \\
w(t) &=& -\frac{\psi^2(t)}{\Delta^2\Omega_-^2}+ g(t)+ 2 i
\Omega_- \int_0^t dt' g(t') .
\end{eqnarray}
Then $x(t)$ in Eq.~(\ref{x:fv}) is given by
\begin{eqnarray}\label{x:mid}
x(t) =- \frac{f v_-(t)}{\Delta \Omega_-^2}\left[1 + i \Omega_- t-
e^{i \psi(t)}\right]\left(1+ \epsilon \frac{t^2}{\Delta^2}\right)
 + \frac{f v_-(t)}{\Delta} \epsilon x_1(t)
 ,
\end{eqnarray}
where
\begin{eqnarray}\label{x_1}
x_1(t) &=& 3 e^{i
\psi(t)}\left[\left(t-\frac{2i}{3\Omega_-}\right) \int^t
\frac{t'^2}{\Delta^2}e^{-i\psi(t')}dt' -\int_0^tdt' \int_0^{t'}
dt'' \frac{t''^2}{\Delta^2}e^{-i\psi(t'')}
\right]   \\
&+& \frac{\Omega_-^2}{v(t)} \left[-t \int_0^t v(t')
 w(t') dt' + \int_0^t dt' \int_0^{t'}dt'' v(t'')
 w(t'')\right] . \nonumber
\end{eqnarray}

In the region $t\geq \Delta$, $r(t)$ can be written as:
\begin{eqnarray}\label{r2:large}
r(t)= \bar{r}[1 +\epsilon g(t)] ,
\end{eqnarray}
where
\begin{eqnarray}\label{s:r}
\epsilon g(t)=\frac{r(t)}{\bar{r}} -1
\end{eqnarray}
is an oscillating function of $O(\epsilon)$ with vanishing
time average and
\begin{eqnarray}\label{r:avg}
\bar{r}=r_{+}\left[1+ \epsilon \frac{\sin^2\Omega_+
\Delta/2}{(\Omega_+ \Delta/2)^2}\right].
\end{eqnarray}
In the adiabatic limit, the inhomogeneous solution
$\rho_{I+}(t)$ vanishes. The phase variable $\phi(t)$ can
be exactly integrated in this case:
\begin{eqnarray} \label{phi:t2}
\phi(t) &=& \phi(\Delta)+ \int_\Delta^t
\frac{L}{r^2(t')}dt' \\
&=&\phi_0- \frac{b}{2\bar{\omega}(r_+^2-b^2)}
  \frac{ \sin\frac{\bar{\omega}(t-t_0)}{2}}{r_+
   + b \cos \frac{\bar{\omega}(t-t_0)}{2}}+
   \frac{1}{\bar{\omega }r_+\sqrt{r_+^2-b^2}}
\tan^{-1}\frac{\sqrt{r_+^2-b^2} \tan
\frac{\bar{\omega}(t-t_0)}{2} }{r_+ +b},  \nonumber \\
&\simeq& \phi_0- \frac{b}{2\bar{\omega}r_1^3}
  \sin\frac{\bar{\omega}(t-t_0)}{2}+
   \frac{1}{\bar{\omega }r_+ ^2}
   \left[\frac{\bar{\omega}(t-t_0)}{2}-\frac{\frac{b}{r_+}
   \tan \frac{\bar{\omega}(t-t_0)}{2}}{1+ \frac{b^2}{r_+^2}
   \tan^2\frac{\bar{\omega}(t-t_0)}{2}}\right], \nonumber\\
&=& \frac{t-t_1}{2 r_+^2} + \mbox{finite periodic
function},
\end{eqnarray}
where $b \cos \bar{\omega}(t-t_0)
=[s_0(\Delta)-(r_1-r_0)]\cos \bar{\omega}(t-\Delta) +
\frac{\dot{s}_0(\Delta)}{\bar{\omega} } \sin
\bar{\omega}(t-\Delta)$ and $\phi_0$, $t_1$ are constants
which are needed to satisfy the boundary condition at
$t=\Delta$. In the case of a non-adiabatic change, we
cannot integrate Eq.~(\ref{phi:t}) exactly but Taylor's
expansion with Eq.~(\ref{r2:large}) gives a good
approximation of Eq.~(\ref{phi:t})up to $O(\epsilon)$:
\begin{eqnarray}\label{phi:approx}
\phi(t) &=& \phi(\Delta)+ \int_\Delta^t
\frac{L}{r^2(t')}dt'  \\
&\simeq& \frac{t-t_1}{2\bar{r}^2} -\epsilon
\frac{\int^t_\Delta dt' g(t')}{\bar{r}^2}, \nonumber
\end{eqnarray}
where $\phi(\Delta)$ is the phase at $t=\Delta$. With these
$\tilde{v}(t)$ is approximated as
\begin{eqnarray}\label{v:t}
\tilde{v}(t) = \left[1 +\epsilon w_+(t) \right] v_+(t)
=\left[1+ \epsilon w_+(t) \right] \bar{r}
e^{-i(t-t_1)/(2\bar{r}^2)} ,
\end{eqnarray}
where
\begin{eqnarray}\label{bars}
v_+(t)&=& \bar{r} e^{-i (t-t_1)/2\bar{r}^2} \\
w_+(t)&=& g(t) +i \frac{\int^t_\Delta g(t')
dt'}{\bar{r}^2}.
\end{eqnarray}
Using
\begin{eqnarray}\label{x:large}
x(t) &=& -\int_0^t f(t')[v_+(t')+\epsilon v_+(t')
w_+(t')]dt' \\
&+& \int_0^\Delta dt' f(t')[v_+(t')+\epsilon v_+(t')
w_+(t') - v_-(t')+ \epsilon v_-(t') w_-(t') ],\nonumber
\end{eqnarray}
one finally obtains
\begin{eqnarray}\label{x:detail}
x(t)&=& -2i \bar{r}^2 f v_+(t) \left[1- \frac{\sin
[\Delta/(4\bar{r}^2)]}{\Delta/(4\bar{r}^2)} e^{i
\frac{2t-\Delta}{4 \bar{r}^2}}- \epsilon y
e^{i\frac{t-\Delta}{2\bar{r}^2}} + \epsilon z(t) \right] ,
\end{eqnarray}
where
\begin{eqnarray}
&& \epsilon y= 1-
      \frac{\sin[\Delta/(4\bar{r}^2)]}{\Delta /(4 \bar{r}^2)}
      e^{i \frac{\Delta}{4 \bar{r}^2} }
- \frac{e^{i(\Delta-t_1)/2\bar{r}^2- i \psi(\Delta)}}{(2
\Omega_+)^{3/2}\bar{r}^3}\left[1
    -\frac{\sin[\psi(\Delta)/2]}{
      \psi(\Delta)/2} e^{i\psi(\Delta)/2}
      \right] ,  \\
&& -2i \bar{r}^2 v_+(t) z(t) =\frac{1 }{\Delta
 \Omega_+^2}v_-(\Delta)x_1(\Delta) -
    \int_\Delta^t v_+(t') w_+(t') dt' .\nonumber
\end{eqnarray}


\end{document}